\documentclass[lettersize,journal]{IEEEtran}

\usepackage{scalerel}
\usepackage{tikz}
\usetikzlibrary{svg.path}

\definecolor{orcidlogocol}{HTML}{A6CE39}
\tikzset{
  orcidlogo/.pic={
    \fill[orcidlogocol] svg{M256,128c0,70.7-57.3,128-128,128C57.3,256,0,198.7,0,128C0,57.3,57.3,0,128,0C198.7,0,256,57.3,256,128z};
    \fill[white] svg{M86.3,186.2H70.9V79.1h15.4v48.4V186.2z}
                 svg{M108.9,79.1h41.6c39.6,0,57,28.3,57,53.6c0,27.5-21.5,53.6-56.8,53.6h-41.8V79.1z M124.3,172.4h24.5c34.9,0,42.9-26.5,42.9-39.7c0-21.5-13.7-39.7-43.7-39.7h-23.7V172.4z}
                 svg{M88.7,56.8c0,5.5-4.5,10.1-10.1,10.1c-5.6,0-10.1-4.6-10.1-10.1c0-5.6,4.5-10.1,10.1-10.1C84.2,46.7,88.7,51.3,88.7,56.8z};
  }
}

\newcommand\orcidicon[1]{\href{https://orcid.org/#1}{\mbox{\scalerel*{
\begin{tikzpicture}[yscale=-1,transform shape]
\pic{orcidlogo};
\end{tikzpicture}
}{|}}}}

\usepackage[utf8]{inputenc} % allow utf-8 input
\usepackage[T1]{fontenc}    % use 8-bit T1 fonts
\usepackage{hyperref}       % hyperlinks
\usepackage{url}            % simple URL typesetting
\usepackage{booktabs}       % professional-quality tables
\usepackage{amsfonts}       % blackboard math symbols
\usepackage{nicefrac}       % compact symbols for 1/2, etc.
\usepackage{microtype}      % microtypography
\usepackage{xcolor}         % colors
\usepackage{color,soul}

\usepackage{array}
 \usepackage{float}
\usepackage{multirow}
\usepackage{amsmath,amssymb,amsfonts}
\usepackage{graphicx}
\usepackage{algpseudocode, algorithm}
\usepackage{caption}
\usepackage{subcaption}

\usepackage{multicol}

\definecolor{benign_blue}{RGB}{102, 102, 255}
\definecolor{mal_red}{RGB}{255,51,51}
\definecolor{newsample_yellow}{RGB}{255,217,102}

\title{Model X-Ray: Detection of Hidden Malware in AI Model Weights using Few Shot Learning}

\author{Daniel Gilkarov \orcidicon{0009-0008-9274-802X},
\IEEEmembership{Student Member, IEEE},\\
Ran Dubin \orcidicon{0000-0002-2055-2211},
\IEEEmembership{Member, IEEE}
\thanks{Daniel Gilkarov is with the Department of Computer Science and Ariel Cyber Innovation Center, Ariel University, Ariel, Israel. e-mail: (daniel.gilkarov1@msmail.ariel.ac.il).}
\thanks{Ran Dubin is with the Department of Computer and Software Engineering and Ariel Cyber Innovation Center, Ariel University, Ariel, Israel. e-mail: (rand@ariel.ac.il).}% <-this % stops a space
\thanks{Manuscript received ????; revised ????}}

\begin{document}

\maketitle

\begin{abstract}
  The potential for exploitation of AI models has increased due to the rapid advancement of Artificial Intelligence (AI) and the widespread use of platforms like Model Zoo for sharing AI models. Attackers can embed malware within AI models through steganographic techniques, taking advantage of the substantial size of these models to conceal malicious data and use it for nefarious purposes, e.g. Remote Code Execution. Ensuring the security of AI models is a burgeoning area of research essential for safeguarding the multitude of organizations and users relying on AI technologies. This study leverages well-studied image few-shot learning techniques by transferring the AI models to the image field using a novel image representation. Applying few-shot learning in this field enables us to create practical models, a feat that previous works lack. Our method addresses critical limitations in state-of-the-art detection techniques that hinder their practicality. This approach reduces the required training dataset size from 40000 models to just 6. Furthermore,  our methods consistently detect delicate attacks of up to 25\% embedding rate and even up to 6\% in some cases, while previous works were only shown to be effective for a 100\%-50\% embedding rate. We employ a strict evaluation strategy to ensure the trained models are generic concerning various factors. In addition, we show that our trained models successfully detect novel spread-spectrum steganography attacks, demonstrating the models' impressive robustness just by learning one type of attack. We open-source our code to support reproducibility and enhance the research in this new field.
\end{abstract}

\section{Introduction}
\label{sec:intro}

\IEEEPARstart{d}{eep} Learning (DL) techniques are commonly used for various tasks such as Computer Vision (CV) \cite{rw_cv_1, rw_cv_2}, Natural Language Processing (NLP) \cite{rw_nlp_1}, Robotics \cite{rw_robo_1, rw_robo_2}, and more. 
One of the most significant advancements in Artificial Intelligence (AI) has been the development of Large Language Models (LLMs). LLMs enabled human-like conversations \cite{chatgpt}, making it easier for people to interact with AI models. This has accelerated the integration of AI models into everyday life, especially among laypeople.
Today, enterprises leverage a diverse set of pre-trained models obtained from model repositories, such as those hosted by Hugging Face \cite{Huggingface} and TensorFlow Hub \cite{tensorflowhub}. These models serve as powerful starting points, allowing organizations to fine-tune them to meet their specific requirements \cite{intro_ft_1} using significantly less data than required for training a model from zero. But, despite the great benefits of model sharing, they also expose end-users to new and untreated cyber threats that can potentially utilize the DL models for malicious purposes using steganography \cite{wang2021evilmodel, wang2022evilmodel, stegonet}, serialization attacks \cite{verma2023insecure}, or both \cite{maleficnet}.

Cyber attackers look for ways to deliver malware to target systems and avoid detection. Among many techniques, 
one common technique they utilize is steganography \cite{img_steg_attack, audio_steg_attack}. Steganography is the art of data hiding, this is a well-researched and developed group of techniques \cite{wang2021data} that have been primarily used and researched inside images \cite{subramanian2021image}, audio \cite{alsabhany2020digital}, video \cite{liu2019video}, and text \cite{majeed2021textreview}. Among many different steganography techniques, Least Significant Bit (LSB) steganography is a prominent variant of steganography \cite{bansal2020survey}. These techniques embed data inside the least significant portions of the cover data, and in doing so, the noticeable effect is minimized.
Similar to many well-known LSB steganography attacks on various digital media types, LSB steganography can be used in DL models. It can be exploited for malicious purposes e.g. for delivering malicious payloads \cite{wang2021evilmodel, wang2022evilmodel, gilkarov2023steganalysis}. However, in contrast, DL models are often much larger, and therefore, they may offer much greater hiding capacity than image, audio, or text data. The increased data transmission capacity may enable an attacker to use attacks that weren't possible before or that are harder to carry out. These techniques take advantage of redundancy, and DL models often use float parameters (32/16 bits), which offer great precision but may sometimes be redundant (see Section \ref{sec:model_performance}). 
To overcome these attacks, several works \cite{10309120,amit2023transpose} suggested ways to disrupt the attacks by modifying the parameters or their representations by using quantization or random bit modification \cite{10309120} or by fine-tuning \cite{amit2023transpose}. 
These methods can possibly be countered by more sophisticated data-hiding techniques that utilize self-correcting codes to resist data corruption. In \cite{maleficnet}, novel AI model steganography techniques utilize low-density parity-check codes, demonstrating resilience to parameter pruning and fine-tuning. Therefore, a detection approach is needed.

Steganalysis techniques are a set of methods aimed at detecting hidden messages in cover data, and they are designed to counter steganography. These detection techniques are an alternative strategy to zero-trust methods that aim to damage the payload by changing the model; instead, the aim is to detect the presence of hidden data. The detection approach has certain benefits over the techniques mentioned earlier: by using detection to filter potentially unwanted files, model repositories or end-users can avoid tampering with the models, which can cause unwanted side effects. In addition, detection may prove effective in cases where the data-hiding methods have some resilience to deliberate data corruption \cite{maleficnet}, as mentioned above. This article explores and develops steganalysis techniques for AI model LSB steganography attacks that can be applied in real-world scenarios. 

One major drawback in traditional ML and DL learning algorithms is that they often require large amounts of training data to avoid overfitting \cite{neyshabur2017exploring}.
This can be problematic for real-life scenarios where we want to detect steganography attacks inside AI models since they vary in size, architecture, task, and many more factors.
The major obstacle we encounter is data scarcity. We require attacked models to research AI model LSB steganography, but no public sources exist. To this end, we build upon the methodology and framework for simulating attacks and creating datasets for research published in previous work \cite{gilkarov2023steganalysis, gilkarov2023steganalysisrepo}. In addition, we focus on DL techniques that are specialized for use in data-scarce scenarios.

One-shot Learning (OSL) and Few-Shot Learning (FSL) are techniques developed to use DL in areas with significantly less training data than traditional DL models require. The core concept is that a model is trained only using one or a few samples per label. OSL and FSL are commonly utilized for CV tasks \cite{intro_osl_cv_1, intro_osl_cv_2, intro_fsl_cv_1}.
There are different approaches to tackling OSL and FSL - augmenting training data to get more samples \cite{intro_fsl_da_1}, fine-tuning pre-trained models \cite{intro_fsl_ft_1}, training embedding models to project embedding samples to a smaller embedding space \cite{intro_fsl_embedding_1}, etc.
A popular FSL approach is called metric-learning \cite{kaya2019deepmetric}. In metric learning, samples are clustered based on their distance from other samples. In a classification context, a sample is labeled based on proximity to other labeled samples according to a certain distance function (like $l_1$ or $l_2$). The widely-known KNN algorithm is a canonical example. Metric learning is often paired with an embedding mechanism that embeds samples to a lower dimension (for example, a convolutional network \cite{Koch2015SiameseNN}). This is called embedding learning. In the training phase, the goal is to minimize the distance between similar samples and maximize the distance between dissimilar samples (in classification, same-label samples are "similar" and vice versa for samples with different labels). 

FSL techniques have been found to have extensive applications in the field of computer vision. Convolutional Neural Networks (CNN) are one of the most popular neural network types used in this domain, and a significant amount of research on FSL involves embedding learning using CNN networks as embedding functions with image data. Some researchers have also transformed data into image representations and applied OSL and FSL techniques to them \cite{HSIAO20191863}.
In this article, we apply FSL techniques to AI models by creating image representations from them, allowing us to apply the rich and deeply developed image FSL techniques to our field. We use metric learning by training a CNN to use as an embedding function and we show that we can effectively classify unseen samples using 2 different evaluation techniques. We compare results and identify different use cases for the different methods.

\subsection{Our contribution}
The key contributions of our research are enumerated below:
\begin{itemize}
\item AI model representation: We propose a method for creating image representations from model weights to transfer problems involving AI models to the well-researched image domain.
\item AI model LSB steganalysis methods: Our research marks the second stepping-stone towards effective and practical techniques for detecting attacks on AI models. The proposed methods have several key benefits that make them practical as opposed to past research:
\begin{itemize}
    \item Big data requirement: Our methods only use up to 6 models for the training phase instead of $\sim$40000 models in the past work \cite{gilkarov2023steganalysis}.
    \item Detection capabilities: Proposed techniques successfully detect attacks up to 25\% Embedding Rate (lower ER is harder) and even 6\% in some cases. Past research \cite{gilkarov2023steganalysis} showed successful results only on 100\%-50\% ER.
    \item Effectiveness on common architectures: Our research shows success on popular large CNN model architectures (for example, ResNet, Densenet, etc.). In contrast, past research techniques \cite{gilkarov2023steganalysis} were only proven effective on small custom CNN architectures ($\sim$2k parameters).
    \item Generality: Using a strict and broad evaluation strategy, we show that our trained models can detect attacks in models that differ from the training data in various properties such as model architecture, size, attack severity, and embedded malware, while past works \cite{gilkarov2023steganalysis} trained and tested models for each combination of those properties individually.
    \item Attack type: We show that models trained using our suggested methodology can detect unknown attacks by training on the simulated LSB steganography attacks.
    
\end{itemize}
%\item Limitations: We conclude the success and drawbacks of our proposed methods and present the future points for improvements.
\item Promotion of reproducible research: To further the development in this nascent field, we have made our steganography attack tools, steganalysis methods, model image representation methods, and the corresponding code publicly available \cite{ourgitrepo}.
\end{itemize}

The remainder of this paper is structured as follows: 
Section \ref{sec:background} provides background knowledge.
Section \ref{Methodology} describes the proposed methodology for data creation, pre-processing, and classification.
Section \ref{sec:experiments} presents the experimental results and analysis.
Section \ref{Limitations} discusses the future research direction and limitations.
Finally, conclusions are drawn in Section \ref{Conclusion}.

\section{Background}
\label{sec:background}
\subsection{X-LSB-Attack}
\label{sec:x_lsb_attack}
X-LSB-Attack \cite{gilkarov2023steganalysis} embeds a malware sample $m$ inside an ordered collection of float numbers. The attack works by using LSB substitution similar to other basic LSB steganography methods in other types of data (images \cite{lsb_steganography_images_1}, audio \cite{lsb_steganography_audio_1}, etc.) The attack embeds $m$ in the $X$ least significant bits (mantissa/fraction part \cite{float32_wiki, float16_wiki}) of the float numbers, similarly to LSB-substitution procedures introduced in previous works \cite{wang2021evilmodel, wang2022evilmodel}.
We use an extra version of X-LSB-Attack called X-LSB-Attack-Fill \cite{gilkarov2023steganalysis} that fills the whole cover data by repeating $m$ or using the largest prefix of it that fits in the $X$ least significant bits of all parameters in the model. While past work \cite{gilkarov2023steganalysis} focused only on float32 data, we generalize the methodology to the other data types. We define $ms$ as the mantissa size of the current data type. For float32, it is 23, and for float16, it is 10. Since changes outside the mantissa are easily noticed \cite{gilkarov2023steganalysis} (see Section \ref{sec:model_performance}), we focus on embedding data in the mantissa regions. See Algorithm \ref{alg:x_lsb_attack} for pseudo-code of \textbf{X-LSB-Attack} and Algorithm \ref{alg:x_lsb_attack_fill} for pseudo-code of \textbf{X-LSB-Attack-Fill}.

\subsection{Embedding Rate}
The embedding rate is a measure for steganography attacks. It measures how much (\%) of the cover data is embedded with hidden data. In the context of steganography inside AI model weights, this can be expressed by 
\[ER = \frac{n_b}{n_w \cdot s}\]
where $n_b$ denotes the number of bits in the embedded payload, $n_w$ denotes the number of weights in the cover model, and $s$ denotes the size of the models' weights (e.g., 32 if the parameters are float32).
Since our experiments use X-LSB-Attack-Fill (Section \ref{sec:x_lsb_attack}) we can express it directly:
\[ER = \frac{n_w\cdot X}{n_w \cdot s} = \frac{X}{s}\]

\section{Methodology}
\label{Methodology}
This section illustrates the methodologies for developing FSL models to detect LSB steganography attacks in AI models.
The main workflow in this article consists of three phases: dataset creation (Section \ref{sec:dataset_creation}), model training (Section \ref{sec:model_train}), and model evaluation (Section \ref{sec:model_test}).
\subsection{Dataset Creation}
\label{sec:dataset_creation}
\begin{figure}[!h]
\centering
\includegraphics[width=1\linewidth]{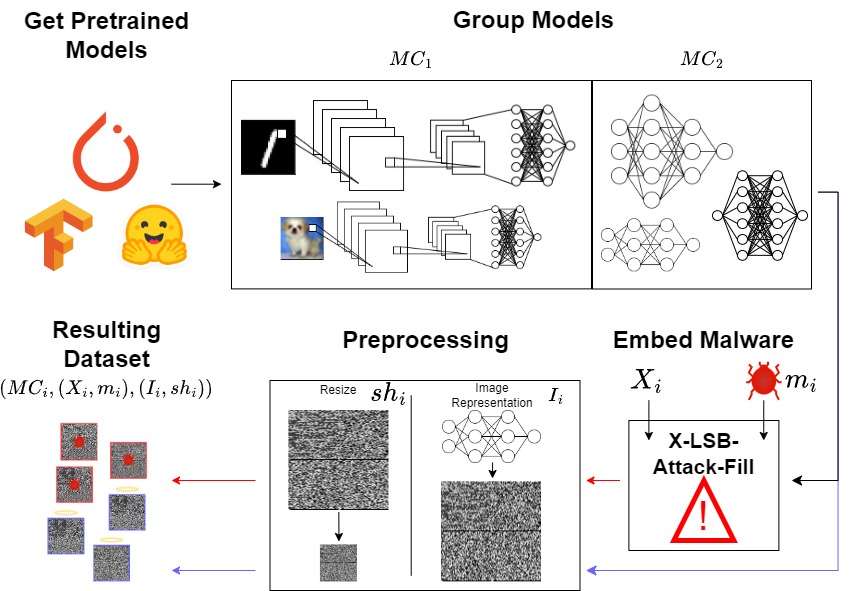}
\caption{Overall dataset creation process (Section \ref{sec:dataset_creation}). First, we get pre-trained models from model repositories and create groups of model zoos based on model size. We call them Model Collections. In this example, we have $MC_1$ with 2 CNNs and $MC_2$ with 3 different DNNs. Second, we use each model collection $MC_i$ to create an attacked version of it by using X-LSB-Attack-Fill (Section \ref{sec:x_lsb_attack}) with some value $X_i$ and malware payload $m_i$. Finally, the original and attacked MCs (colored \textcolor{benign_blue}{blue} and \textcolor{mal_red}{red} respectively) go through a pre-processing phase. In this example, we calculate some model image representation $I_i$ (Section \ref{sec:image_creation}) and reshape all resulting images to shape $sh_i$. The process results in 2 datasets of \textcolor{benign_blue}{benign} and \textcolor{mal_red}{malicious} model image representations created from $MC_1$ and $MC_2$.}
\label{fig:dataset_creation}
\end{figure}

Datasets for training FSL models are created by following 3 steps: model collection, embedding malware, 
% Feature extraction (optional),
and pre-processing.
Every model collection is processed into a binary classification dataset of benign (labeled 0) and malicious models (labeled 1). See Figure \ref{fig:dataset_creation} for an illustration.

\paragraph{Model Collection}
\label{sec:model_collection}
Model Zoos (MZ) are a population of models of the same architecture trained on the same task and can have only one model. We define a Model Collection (MC) as a collection of MZs. MCs for FSL are created by downloading pre-trained popular models from online model hubs like Tensorflow-Hub \cite{tensorflowhub} and grouping them based on their size. 

\paragraph{Embedding Malware}
\label{sec:embedding_malware}
Using the MC of pre-trained benign models, we create attacked versions of it by attacking every model in the collection using the same attack.
We use the X-LSB-Attack-Fill method (Section \ref{sec:x_lsb_attack}) to attack the models using a malware sample $m$ from MalwareBazaar \cite{MalwareBazaar}. We embed multiple malware payloads or randomly generated binary payloads for large models (>300M parameters). The float mantissa region has $ms$ bits, so we use X-LSB-Attack-Fill with $1 \le X \le ms$ - this results in $ms$ different attacked model collections of different attack severity, a lesser $X$ value means a more subtle, harder-to-detect attack. 

\paragraph{Pre-Processing}
\label{sec:pre_processing}
We apply pre-processing to our data to make it more suitable for our FSL models (siamese CNNs that work with constant-sized images). We developed techniques for creating images from float numbers (model parameters, etc.). See section \ref{sec:image_creation} for an in-depth explanation of the procedure. After we convert our data into image form (8-bit grayscale), we reshape (downscale or upscale) the images into a constant predetermined size and normalize the 0-255 integer values to the 0-1 real values range - this is common practice for training CNNs \cite{7808140}.

\paragraph{Dataset Hyperparameters}
\label{sec:dataset_hyperparameters}
Using our methodology, datasets can be created in a large variety of ways. Every dataset $D$ will have a set of hyperparameters $hp = (MC, (X, m), (I, sh))$ that influenced its creation. Each step in the creation procedure uses the hyperparameters. $MC$ is the Model Collection of MZs from which the dataset was created. $(X, m)$ are used in the embedding procedure - $X$ is the amount of LSBs we chose to override and $m$ is the malware sample that was embedded. 
% $F$ is the feature that was extracted. 
$(I, sh)$ are used in the pre-processing phase, $I$ is the image transformation used (See \ref{sec:image_creation}) and $sh$ is the final shape all images were reshaped into.

\subsection{Train/Test Split}
We split the processed image datasets into train and test sets using the following methodology: 
\[Let \; MZS=\{MZ_1, MZ_2, ...\}\]
The model zoos from which we created our datasets. Each model zoo has models of the same architecture trained on the same data task. We pick some subset of the model zoos and set:
\[MZS_{train} \subseteq MZS\]
\[MZS_{test} := MZS \setminus MZS_{train} \]
We compile the training dataset by taking all models (benign and attacked) of the model zoos in $MZS_{train}$, and the test dataset is compiled similarly with $MZS_{test}$. The split is deliberately performed to help prove the trained model's robustness by testing it against models of architectures that weren't seen during training.

\subsection{AI Model Image Representation}
\label{sec:image_creation}
This section will discuss our methods for creating image representations from float model weights. We introduce a novel image representation called Grayscale-Fourpart.

\label{sec:grayscale_fourpart}
\noindent The Grayscale-Fourpart image representation works on float32 data specifically. It is created by dividing each 32-bit model weight $w:=b_{31} \cdots b_{0}$ to 4 8-bit parts: $p_0:=b_{31} \cdots b_{24}, p_1:=b_{23} \cdots b_{16}$ and so on. Each 8-bit long binary string (byte) is used as an unsigned int8 value with values in the range 0-255 like standard 8-bit images and serves as one pixel. 4 images are created by stacking the model weight parts $p_1, p_2, p_3, p_4$ from all weights (all $p_1$'s, all $p_2$'s, ...) and shaping the stacked vector of byte values into square shapes (adding zero padding if needed). 
The 4 images are stacked side to side to form the final image.
See Algorithm \ref{alg:grayscale_fourpart} and Figure \ref{fig:model2img}.

\subsection{Convolutional Embedding Network}
\label{sec:model_train}
This section outlines our methodology for training a CNN to learn embeddings.
We used the triplet method \cite{schroff2015facenet} with $l_2$ distance for the training strategy. We take our train data and construct all possible (anchor, positive, negative) triplets, which in our case is based on the labels (i.e. (benign, benign, malicious) and (malicious, malicious, benign)). We train a CNN feature extractor using Adam \cite{kingma2017adam}. We experiment with three different training strategies: \begin{enumerate}
    \item Early Stopping (ES): train for one epoch.
    \item Standard Training (ST): train for five epochs.
    \item Until Below (UB): train for 100 epochs, and stop training if train loss is within some interval, e.g. [0.5, 1.25]. 
\end{enumerate}

\begin{figure*}[!h]
\centering
\includegraphics[width=\linewidth]{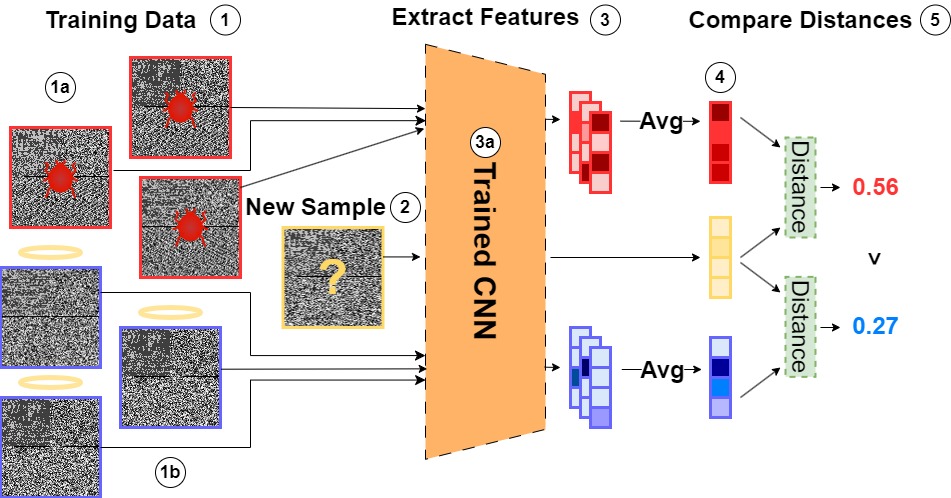}
\caption{Illustration of classifying a new sample (colored \textcolor{newsample_yellow}{yellow}, see (2)) using the trained CNN (3a) with the centroid method. We compute the embeddings of our benign and malicious training samples (colored \textcolor{benign_blue}{blue} and \textcolor{mal_red}{red}, see (1a) and (1b)) and average them to get the centroid embeddings (4). Then, we classify the new sample by measuring the $l_2$ distance of its embedding to both centroid embeddings and giving it the label for the closer of the two (5). In this example, the \textcolor{newsample_yellow}{new sample} is labeled \textcolor{benign_blue}{benign} because it got 0.27 distance to the benign centroid as opposed to 0.56. This procedure is essentially the same as applying 1-Nearest-Neighbor w/$l_2$ distance using the centroids' embeddings.}
\label{fig:eval_centroids}
\end{figure*}

\label{sec:model_test}
\noindent Testing phase: Model evaluation is done with 2 different procedures that show competitive results. We focus on methods that can be practically used.
\paragraph{Centroid} We classify a new sample using the centroids of the benign and malicious training data. This is common practice in FSL \cite{wang2019simpleshot}. Let $D_{0}$ and $D_{1}$ denote them, let $I$ denote the image to be labeled, and $f$ to be the trained CNN. A centroid of some data $D_i$ is given by $c_i= \frac{1}{|D_i|}\sum_{d \in D_i} f(d)$. $I$ is then given the label $argmin_{i \in \{0,1\}} \; l_2(f(I), c_i)$.
See Figure \ref{fig:eval_centroids}.

\paragraph{KNN} We classify a new sample using a similar process to the centroid technique, but we apply KNN using all embeddings of the benign and malicious training data instead of the average embeddings. 

\section{Experiments}
\label{sec:experiments}
In this section, the experimental results are laid out and analyzed. The experiments are carried out by exercising the data creation and classification methodology discussed in Section \ref{Methodology}. In each experiment, we tried out different MCs that we had compiled. We train models using some dataset $D$. We evaluate them on all attack severities ($1 \le X \le s$) in the same MC (in-distribution) and other MCs (out-of-distribution) using the test data that contains unseen model architectures to analyze how robust they are with regard to model architecture, embedded malware, model size, model task, and attack severity. More specifically, we aim to show that the trained models can be used in real-life scenarios. All experiments were run on an NVIDIA GeForce RTX 3080 GPU, i9 intel core, and 32GB RAM.

\subsection{Baseline}
\label{sec:baseline}
Research on detecting attacks on AI models is a relatively new field in its early stages and has few existing works \cite{verma2023insecure,gilkarov2023steganalysis}. Recent work \cite{verma2023insecure} proposed detecting malicious model serialization attacks, which is outside the scope of our work. 
The work done in \cite{gilkarov2023steganalysis} focused on detecting LSB steganography in AI models. It is the closest to the scope of this work. Hence, we use it as a baseline.
The baseline work \cite{gilkarov2023steganalysis} trains unsupervised and supervised models using the small CNN zoos we introduced in Section \ref{sec:datasets}, and it showed non-trivial results for X $\ge$ 17 on the CIFAR10, MNIST, SVHN, and STL10 MZs individually. We closely follow the evaluation process used in those experiments to create a fair comparison between our methods and the baseline.

\subsection{Chosen Feature Extraction CNNs}
We use two CNN architectures as feature extraction for our FSL models (See Section \ref{sec:model_train}):

\noindent \textbf{OSL CNN}:
We utilize the well-established CNN architecture designed for OSL introduced in \cite{koch2015siamese} with images resized to 100x100. We denote this architecture as \textbf{OSL CNN}.

\noindent \textbf{SRNet}:
Several works develop CNNs focused on image steganalysis. We utilize a prominent residual CNN called SRNet \cite{srnet} in our experiments with images resized to 256x256. We denote this architecture as \textbf{SRNet}.

\subsection{Datasets}
\label{sec:datasets}
We create 4 different datasets for training and testing. We aim to comprehensively analyze different model architectures, sizes, etc. In addition, we reproduce results that use a spread-spectrum attack in the AI models \cite{maleficnet} as a special study case on how the solution can handle unknown steganography attacks. We used the Grayscale-Fourpart image representation (Section \ref{sec:grayscale_fourpart}) with image sizes 100 and 256 for all float32 datasets.

\subsubsection{Small CNN Zoos}
\label{sec:scz}
Following the baseline work, 
We compile a dataset from a model zoo of small CNN models (2k float32 parameters) of the same architecture that was trained on STL10 \cite{schürholt2022model}. We denote this dataset as \textbf{SCZ}. The model zoo has about 50000 different models.
We used the malware sample with sha256 6054f328f8d54d0a54f5e3b90cff020e139105
eb5aa5a3be52c29dbea6289c30 on this dataset (denoted $m_{6054f}$).
Since this dataset is used for comparison with the baseline \cite{gilkarov2023steganalysis} (see \ref{sec:baseline}), we use a train/test split on models from the same model zoo in contrast with our methodology. We use 3 benign and 3 malicious models in the training split and the rest for testing.

\subsubsection{Famous CNNs (Small)}
We compile a MC of different well-known CNN architectures of size at most 10M weights with float32 parameters. We use the imagenet pre-trained models supplied in the Keras library \cite{keras_applications}. The chosen architectures are: DenseNet121 \cite{huang2018densely}, EfficientNetV2 (B0,B1) \cite{tan2021efficientnetv2}, MobileNet (V1, V2, V3Small, V3Large) \cite{howard2017mobilenets, sandler2019mobilenetv2, howard2019searching}, and NASNetMobile \cite{zoph2018learning}.
We used the malware sample with sha256 77e05b52f51cfc8ec31f0dc2e544dc21b94250f
35a5a353fd5e4e271e75bc45d on this model collection (denoted $m_{77e05}$). We chose the MZs MobileNet, NASNetMobile, and MobileNetV3Large for training and the rest for testing.

\subsubsection{Famous CNNs (Large)}
We compile an MC of well-known CNN architectures of size at most 100M and more than 10M weights with float32 parameters. We use the imagenet pre-trained models supplied in the Keras library \cite{keras_applications}. The chosen architectures are: \begin{itemize}
    % \item ConvNeXt (Base, Tiny, Small) \cite{liu2022convnet}
    \item DenseNet (169, 201) \cite{huang2018densely}
    \item EfficientNetV2 (B2,B3, S, M) \cite{tan2021efficientnetv2}
    \item Inception (V3, ResNetV2) \cite{szegedy2015rethinking, szegedy2016inceptionv4}
    \item NASNetLarge \cite{zoph2018learning}
    \item ResNet (50(V1, V2), 101(V1, V2), 152(V1, V2)) \cite{he2015deep, he2016identity}
    \item Xception \cite{chollet2017xception}
\end{itemize}
For this MC, we used 3 concatenated malware samples. Their MD5 hashes are: \begin{itemize}
    \item 9de9993c77412ba8fa3200714dcd58f6
    \item edd73e79cb7b8abe0f569844931f4fdb
    \item 6d5c304bef8b2c32e904946f03b601d9
\end{itemize}
Denote this malicious payload $m_{9d\_ed\_6d}$.
We chose the MZs DenseNet169, and NASNetLarge for training and the rest for testing.

\subsubsection{Maleficnet OOD Attack}
\label{sec:maleficnet}
We use the code from the authors in \cite{maleficnet} to recreate spread-spectrum attacks. Using our methodology (Section \ref{Methodology}), we create datasets from the benign and attacked models for additional evaluation. The recreated attacks (architecture + malware payload) are: 
\begin{itemize}
    \item DenseNet121 (Stuxnet, Destover)
    \item Resnet50 (Stuxnet, Destover, Asprox, Bladabindi)
    \item Resnet101 (Stuxnet, Destover, Asprox, Bladabindi, Cerber, Eq.Drug, Kovter)
\end{itemize}
The datasets we created above have 1 benign sample in each MZ and 2/4/7 malicious samples in the DenseNet121/Resnet50/Resnet101 architectures, respectively. Hence, a meaningful classification accuracy (better than weighted random guessing) will be more than 66\%/80\%/87.5\%.

In the context of this paper, we call the dataset used in training an In-Distribution (ID) dataset, and other datasets not used in training are called Out-Of-Distribution (OOD). Studying the model performance on OOD data is essential to analyzing its generability.

\subsection{Error Bars}
\label{sec:error_bars}
When applicable, we perform repeated experiments and show confidence interval plots using the seaborn python package \cite{Waskom2021}. The randomness in the trials stems from the CNN models' random initialization, and train sample shuffling.

\subsection{Model Evaluation Metric}
\label{sec:weighted_metric}
We define a simple metric for grading models. The metric is a weighted accuracy of the model's result on some datasets on all $s$ attack severities and benign samples. We emphasize that attacking the least significant bytes in the mantissa will be much harder to detect (See Section \ref{sec:model_performance}) therefore, we give higher weights to the more complicated detection. Let $a_i$ the accuracy of a model on the dataset with $i$ bits attacked, and let $a_0$ the accuracy of the same model on the benign dataset. The weighted metric is given by:
\[WM = \frac{1}{2} \cdot (a_0 + \sum_{i=1}^{s}\frac{(s-i+1)a_i}{\sum_{i=1}^{s}i})\]
The Weighted Metric is bounded in the range [0,1] by definition.

\subsection{Plot Types}
\label{sec:plot_types}
We use 2 main types of plots to showcase our results:
\begin{enumerate}
    \item Only Model LSB (OML): These plots show classification accuracy of models trained on some attack severity $1 \le \widehat{X} \le 23$ that are then tested on benign and attacked models that were attacked with attack severity $\widehat{X}$. The x-axis shows Model LSB and the y-axis shows mean accuracy for both benign and attacked models. These plots help us get an initial sense of the success of a model.
    \item All LSBs (AL): These plots show the weighted metric (see Section \ref{sec:weighted_metric}) of models trained on some attack severity $1 \le \widehat{X} \le 23$. The x-axis shows Model LSB, and the y-axis shows the weighted metric. These plots help us decide what $X$ value results in the best global-performing model on a specific dataset.
\end{enumerate}

\subsection{Results}
\label{sec:results}
\subsubsection{Experiment 1 - SCZ Baseline}

\begin{figure}
    \centering
    \includegraphics[width=\linewidth]{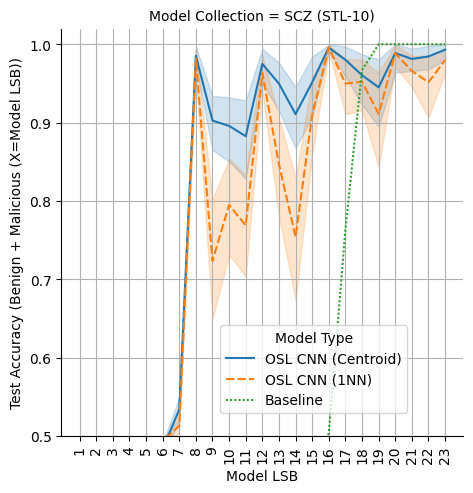}
    \caption{Experiment 1 OML results (see Section \ref{sec:plot_types}). We train OSL CNN models on 6 samples from the SCZ dataset. The baseline result from prior work \cite{gilkarov2023steganalysis} is plotted as a dotted green line.}
    \label{fig:exp1}
\end{figure}

In this experiment, we train classification models on the SCZ MC (Section \ref{sec:scz}) using the UB training strategy (see Section \ref{sec:model_train}). The main goal of this experiment is to measure the success (test accuracy) of models trained using our proposed methodology (Section \ref{Methodology}) versus the baseline results \cite{gilkarov2023steganalysis}. To this end, we closely replicate the training/testing procedure they used to make the comparison strict. The baseline work trained a model for each attack severity and each model architecture individually. Here, we focus on the STL10 model zoo and train our classification models on 3 STL10 models (benign and attacked) on each attack severity (23 different models). We repeat the training process 30 times and show mean and confidence interval results. See OML Figure \ref{fig:exp1}. The plot shows the \#LSB in the training dataset each model learned versus the test accuracy on the STL10 MZ. The CNN model with centroid evaluation had very close results to the baseline for $\#LSB \in [18,23]$ (56\%-71\% ER). For $\#LSB \in [8,18]$ attacks (25\%-56\% ER), it showed stable results in the confidence interval [0.85, 1], while the baseline model failed. The CNN model with 1NN evaluation also had good results but was less stable. This is likely due to the model being overfit. In conclusion, we successfully trained FSL models that outperformed the baseline in terms of accuracy using only 6 models for training, as opposed to 40,000 models in the baseline method.

\subsubsection{Experiment 2 - small famous CNNs}
In this experiment, we train classification models on the small famous CNNs using the UB training strategy (see Section \ref{sec:model_train}). The main goal of this experiment is to see how the techniques fare with the larger CNN architectures and whether the models succeed on unseen model architectures and unseen attacks. We train models on all 23 attack severities like before, but this time, we measure their performance on the other attacks (unseen during training) using the  Weighted Metric introduced in Section \ref{sec:weighted_metric}. This metric gives us a general way to score models and rewards success on the more delicate attacks (smaller $X$ value), which are a major challenge in this area. In addition to evaluating the models on the small famous CNNs test set, we also evaluate the models on the large famous CNNs (See Section \ref{sec:datasets}). These architectures serve as an OOD test for the models trained on the small famous CNNs, with a size difference of (10M-100M). We show average results over 50 repeated trials and confidence intervals. We start by examining AL and OML results on the small CNNs test set.

\begin{figure}
    \centering
    \includegraphics[width=\linewidth]{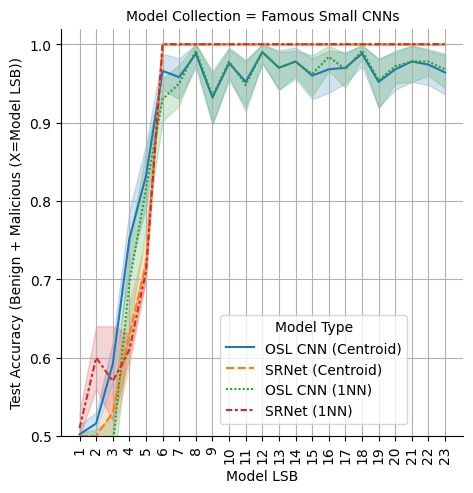}
    \caption{Experiment 2 OML ID results. FSL models trained on the famous small CNNs train set and tested on the small CNNs test set.}
    \label{fig:exp2_oml_id}
\end{figure}

From OML Figure \ref{fig:exp2_oml_id}, we can see that SRNet models (centroid and NN evaluation) from $X \ge 6$ got perfect accuracy with great confidence, and OSL CNN models were close behind with accuracy in the confidence interval $[0.9, 1]$. These are substantial results with good accuracy for ERs 19\%-100\% as opposed to 50\%-100\% in the baseline work. As we gather from Section \ref{sec:model_performance}, attacks with ER 3\%-50\% largely don't affect model performance which means such attacks are likely to be used, and so we need effective countermeasures for them. SRNet with 1NN evaluation had remarkable results in attacks with $X=2$ (ER=3.1\%) with a confidence interval of approximately 0.55-0.65. OSL CNN models with centroid evaluation were the most successful for $X \in \{3, 4,5\}$ with mean accuracy 0.6, 0.7, and 0.8 respectively with good confidence. After examining the models' performance on the attack severity they trained on, we move on to examine the results more globally on all attack severities.

\begin{figure}
    \centering
    \includegraphics[width=\linewidth]{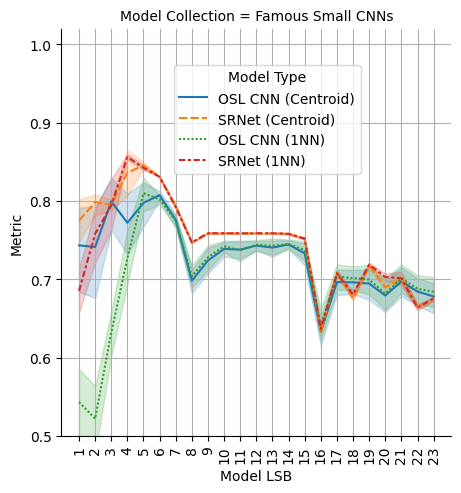}
    \caption{Experiment 2 AL ID results. FSL models trained on the famous small CNNs train set and tested on the small CNNs test set.}
    \label{fig:exp2}
\end{figure}

From AL Figure \ref{fig:exp2}, we see models trained on subtle attacks had the highest success globally. The best weighted metric result is $\sim$85\% using SRNet on $X=4$. This is an impressive result since the metric, by definition, requires adequate performance on both benign and malicious data. In particular, with a metric rating of 0.85, accuracy on benign data must be $\ge 70\%$. Concluding the AL results, we see that models trained using our proposed methodology achieve good results with very few training samples, and we prove their worth for all attack ranges as opposed to the prior work, which trained a separate model for each attack severity.

\begin{figure}
    \centering
    \includegraphics[width=\linewidth]{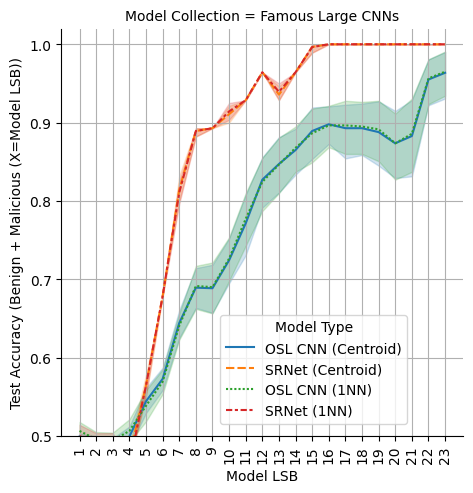}
    \caption{Experiment 2 OML OOD results. FSL models trained on the famous small CNNs train set and tested on the large CNNs test set.}
    \label{fig:exp2_oml}
\end{figure}

Moving on, we look to evaluate the models on OOD data.
From OML Figure \ref{fig:exp2_oml}, which shows results on the large famous CNNs, we can see the models exhibit accuracy $\ge 70\%$ for $X \ge 6$ and $X \ge 9$ with the SRNet and OSL CNN models respectively. There's no noticeable difference between the centroid and 1NN evaluation. We note the SRNet models have perfect accuracy for $X \ge 16$ ($\ge 50\%$ ER) with great confidence.
We feel these results are a good demonstration of the models' generality with regard to model architecture and embedded payload.

\subsubsection{Experiment 2.5 - MaleficNet}

We take the models trained in Experiment 2 and evaluate them on the MaleficNet attacked models presented in Section \ref{sec:datasets}. See Table \ref{tab:malefic}. We display the Run Num where the model was created, the model LSB, evaluation type, weighted metric on ID data, and mean accuracy on maleficnet data. We filter models which had $\ge60\%$ accuracy on the benign and malicious maleficnet models with 1NN evaluation or with centroid evaluation. Out of these, we show the top 3 with centroid and the top 3 with 1NN evaluations sorted by the Weighted Metric on the ID small famous CNNs test set. We first note that only OSL CNN models had success on MaleficNet data, this shows an added benefit of the OSL model over the SRNet model which had better performance on ID data. We see all models that had success trained on $X \in \{1,2,3\}$, this might be due to the model being fit to detect very delicate changes. The top model (centroid evaluation with $X=1$) in the table exhibited a weighted metric of 87\% and 80\% accuracy on the MaleficNet data, this shows that indeed, some models trained on our unrelated LSB substitution attacks can detect very different spread-spectrum attacks, and simultaneously not compromise ID performance which is substantial. It's important to note that the WM value of 87\% doesn't lie within the confidence interval seen in Figure \ref{fig:exp2} which is approximately (0.75, 0.8) for the OSL CNN models with centroid evaluation. Nevertheless, this is a motivating result. It's also worth mentioning that the MaleficNet models are of DenseNet121, ResNet50, and ResNet101 architectures, of which none are in the small famous CNNs train set (see Section \ref{sec:datasets}). In conclusion, this experiment further outlines the generality of trained models using our methodology, in particular, regarding attack type, embedded payload, and model architecture.

\begin{table}[]
\caption{Classification accuracy of selected models from Experiment 2 on MaleficNet OOD attack}
\label{tab:malefic}
\centering
\begin{tabular}{@{}ccccc@{}}
\toprule
Run Num & Model LSB & Eval Type & WM & OOD Acc \\ \midrule
48      & 1         & C         & \textbf{87} & \textbf{80}      \\
37      & 2         & C         & 82 & 71      \\
31      & 2         & C         & 81 & 80      \\
13      & 1         & NN        & 80 & 64      \\
38      & 1         & NN        & 73 & 67      \\
42      & 3         & NN        & 64 & 79      \\ \bottomrule
\end{tabular}
\end{table}

\section{Limitations and Future Work}
\label{Limitations}
Our research focused on detecting LSB substitution steganography attacks and introduced new steganalysis methods. While experimental results show promising results in LSB attacks and spread-spectrum attacks, future research should cover evasive attack vectors like maleficnet \cite{maleficnet} directly. This work introduces a general methodology and software framework for research \cite{ourgitrepo}, however, we focus on CNN models and particularly float32 data. Future research should investigate expanding the study to include more compact data types, such as 16-bit, 8-bit, 4-bit, or even 2-bit quantization, which are increasingly utilized in various models, including text models. Float16 data can be addressed in our methodology by using a different image representation, for example, creating an image from the last byte might be sufficient as the float16 mantissa is 10 bits.

% While there is a need to explore additional steganography attacks, it is significant to note that the proposed solution was able to identify unknown spread spectrum attacks that were not included in the training sets, indicating its versatility. The research goal was to minimize the samples required to detect steganography attacks accurately. However, a limitation of the current work is the necessity to segregate the FSL model based on architecture size. Future research should explore alternative solutions to address this limitation and consider employing anomaly detection as a possible solution.

\section{Conclusion}
\label{Conclusion}
 In this work, we propose novel few-shot learning methods to detect AI model steganography attacks. These attacks leverage the AI model-sharing community for malicious purposes by hiding malware in the shared models. Protecting end-users from unknowingly downloading malicious content is a fierce challenge. Our experimentation achieves very high detection rates using just a small training dataset that contained up to 6 model weights (samples) compared to 40000 samples of pre-trained models needed by the current state of the art. Furthermore, our methods consistently detect delicate attacks of up to 25\% embedding rate and even up to 6\% in some cases. At the same time, the previous state-of-the-art was only shown to be effective for 100\%-50\% embedding rate. Our experimental results show methods trained using our methodology can even detect OOD novel attacks \cite{maleficnet} that current state-of-the-art prevention techniques don't prevent.

\section{Acknowledgement}
 This work was supported by the Ariel Cyber Innovation Center in conjunction with the Israel National Cyber Directorate in the Prime Minister's Office. This work is under US Provisional Patent Application No. 63/524,681.

% \newpage
\bibliography{main}

\begin{thebibliography}{10}

\bibitem{alsabhany2020digital}
Ahmed~A AlSabhany, Ahmed~Hussain Ali, Farida Ridzuan, AH~Azni, and Mohd~Rosmadi Mokhtar.
\newblock Digital audio steganography: Systematic review, classification, and analysis of the current state of the art.
\newblock {\em Computer Science Review}, 38:100316, 2020.

\bibitem{amit2023transpose}
Guy Amit, Mosh Levy, and Yisroel Mirsky.
\newblock Transpose attack: Stealing datasets with bidirectional training, 2023.

\bibitem{bansal2020survey}
Kriti Bansal, Aman Agrawal, and Nency Bansal.
\newblock A survey on steganography using least significant bit (lsb) embedding approach.
\newblock In {\em 2020 4th International Conference on Trends in Electronics and Informatics (ICOEI)(48184)}, pages 64--69. IEEE, 2020.

\bibitem{intro_fsl_cv_1}
Peyman Bateni, Raghav Goyal, Vaden Masrani, Frank Wood, and Leonid Sigal.
\newblock Improved few-shot visual classification.
\newblock In {\em Proceedings of the IEEE/CVF Conference on Computer Vision and Pattern Recognition (CVPR)}, June 2020.

\bibitem{lsb_steganography_audio_1}
Debnath Bhattacharyya, Poulami Dutta, Maricel~O. Balitanas, Tai-hoon Kim, and Purnendu Das.
\newblock Hiding data in audio signal.
\newblock In Chin-Chen Chang, Thanos Vasilakos, Purnendu Das, Tai-hoon Kim, Byeong-Ho Kang, and Muhammad Khurram~Khan, editors, {\em Advanced Communication and Networking}, pages 23--29, Berlin, Heidelberg, 2010. Springer Berlin Heidelberg.

\bibitem{srnet}
Mehdi Boroumand, Mo~Chen, and Jessica Fridrich.
\newblock Deep residual network for steganalysis of digital images.
\newblock {\em IEEE Transactions on Information Forensics and Security}, 14(5):1181--1193, 2019.

\bibitem{chollet2017xception}
François Chollet.
\newblock Xception: Deep learning with depthwise separable convolutions, 2017.

\bibitem{audio_steg_attack}
Catalin Cimpanu.
\newblock Wav audio files are now being used to hide malicious code.
\newblock \url{https://www.zdnet.com/article/wav-audio-files-are-now-being-used-to-hide-malicious-code/}, 2019.

\bibitem{img_steg_attack}
CyberTalk.
\newblock Hackers hide malware in windows logo.
\newblock \url{https://www.cybertalk.org/2022/09/30/hackers-hide-malware-in-windows-logo/}, 2022.

\bibitem{rw_nlp_1}
Jacob Devlin, Ming-Wei Chang, Kenton Lee, and Kristina Toutanova.
\newblock Bert: Pre-training of deep bidirectional transformers for language understanding, 2019.

\bibitem{10309120}
Ran Dubin.
\newblock Disarming attacks inside neural network models.
\newblock {\em IEEE Access}, 11:124295--124303, 2023.

\bibitem{intro_osl_cv_1}
Li~Fei-Fei, R.~Fergus, and P.~Perona.
\newblock One-shot learning of object categories.
\newblock {\em IEEE Transactions on Pattern Analysis and Machine Intelligence}, 28(4):594--611, 2006.

\bibitem{gilkarov2023steganalysisrepo}
Daniel Gilkarov and Ran Dubin.
\newblock Ai model steganalysis.
\newblock \url{https://github.com/ArielCyber/AI_Model_Steganalysis}, 2023.

\bibitem{ourgitrepo}
Daniel Gilkarov and Ran Dubin.
\newblock Modelxray code repository.
\newblock \url{https://github.com/danigil/ModelXRay.git}, 2024.

\bibitem{gilkarov2023steganalysis}
Daniel Gilkarov and Ran Dubin.
\newblock Steganalysis of ai models lsb attacks.
\newblock {\em IEEE Transactions on Information Forensics and Security}, 19:4767--4779, 2024.

\bibitem{he2015deep}
Kaiming He, Xiangyu Zhang, Shaoqing Ren, and Jian Sun.
\newblock Deep residual learning for image recognition, 2015.

\bibitem{he2016identity}
Kaiming He, Xiangyu Zhang, Shaoqing Ren, and Jian Sun.
\newblock Identity mappings in deep residual networks, 2016.

\bibitem{maleficnet}
Dorjan Hitaj, Giulio Pagnotta, Briland Hitaj, Luigi~V. Mancini, and Fernando Perez-Cruz.
\newblock Maleficnet: Hiding malware into deep neural networks using spread-spectrum channel coding.
\newblock In {\em Computer Security – ESORICS 2022: 27th European Symposium on Research in Computer Security, Copenhagen, Denmark, September 26–30, 2022, Proceedings, Part III}, page 425–444, Berlin, Heidelberg, 2022. Springer-Verlag.

\bibitem{howard2019searching}
Andrew Howard, Mark Sandler, Grace Chu, Liang-Chieh Chen, Bo~Chen, Mingxing Tan, Weijun Wang, Yukun Zhu, Ruoming Pang, Vijay Vasudevan, Quoc~V. Le, and Hartwig Adam.
\newblock Searching for mobilenetv3, 2019.

\bibitem{howard2017mobilenets}
Andrew~G. Howard, Menglong Zhu, Bo~Chen, Dmitry Kalenichenko, Weijun Wang, Tobias Weyand, Marco Andreetto, and Hartwig Adam.
\newblock Mobilenets: Efficient convolutional neural networks for mobile vision applications, 2017.

\bibitem{HSIAO20191863}
Shou-Ching Hsiao, Da-Yu Kao, Zi-Yuan Liu, and Raylin Tso.
\newblock Malware image classification using one-shot learning with siamese networks.
\newblock {\em Procedia Computer Science}, 159:1863--1871, 2019.
\newblock Knowledge-Based and Intelligent Information \& Engineering Systems: Proceedings of the 23rd International Conference KES2019.

\bibitem{huang2018densely}
Gao Huang, Zhuang Liu, Laurens van~der Maaten, and Kilian~Q. Weinberger.
\newblock Densely connected convolutional networks, 2018.

\bibitem{Huggingface}
HuggingFace.
\newblock {Fickling pickle scanning}, 2022.
\newblock Accessed: 2022-12-19.

\bibitem{intro_ft_1}
Ahmadreza Jeddi, Mohammad~Javad Shafiee, and Alexander Wong.
\newblock A simple fine-tuning is all you need: Towards robust deep learning via adversarial fine-tuning.
\newblock {\em arXiv preprint arXiv:2012.13628}, 2020.

\bibitem{lsb_steganography_images_1}
Neil~F. Johnson and Sushil Jajodia.
\newblock Exploring steganography: Seeing the unseen.
\newblock {\em Computer}, 31(2):26--34, 1998.

\bibitem{kaya2019deepmetric}
Mahmut Kaya and Hasan~{\c{S}}akir Bilge.
\newblock Deep metric learning: A survey.
\newblock {\em Symmetry}, 11(9):1066, 2019.

\bibitem{keras_applications}
Keras.
\newblock Keras applications, 2024.
\newblock Accessed: 2024-04-13.

\bibitem{kingma2017adam}
Diederik~P. Kingma and Jimmy Ba.
\newblock Adam: A method for stochastic optimization, 2017.

\bibitem{Koch2015SiameseNN}
Gregory Koch, Richard Zemel, and Ruslan Salakhutdinov.
\newblock Siamese neural networks for one-shot image recognition.
\newblock 2015.

\bibitem{koch2015siamese}
Gregory Koch, Richard Zemel, Ruslan Salakhutdinov, et~al.
\newblock Siamese neural networks for one-shot image recognition.
\newblock In {\em ICML deep learning workshop}, volume~2. Lille, 2015.

\bibitem{intro_fsl_embedding_1}
Chen Liu, Yanwei Fu, Chengming Xu, Siqian Yang, Jilin Li, Chengjie Wang, and Li~Zhang.
\newblock Learning a few-shot embedding model with contrastive learning.
\newblock {\em Proceedings of the AAAI Conference on Artificial Intelligence}, 35(10):8635--8643, May 2021.

\bibitem{stegonet}
Tao Liu, Zihao Liu, Qi~Liu, Wujie Wen, Wenyao Xu, and Ming Li.
\newblock Stegonet: Turn deep neural network into a stegomalware.
\newblock In {\em Annual Computer Security Applications Conference}, ACSAC '20, page 928–938, New York, NY, USA, 2020. Association for Computing Machinery.

\bibitem{liu2019video}
Yunxia Liu, Shuyang Liu, Yonghao Wang, Hongguo Zhao, and Si~Liu.
\newblock Video steganography: A review.
\newblock {\em Neurocomputing}, 335:238--250, 2019.

\bibitem{majeed2021textreview}
Mohammed~Abdul Majeed, Rossilawati Sulaiman, Zarina Shukur, and Mohammad~Kamrul Hasan.
\newblock A review on text steganography techniques.
\newblock {\em Mathematics}, 9(21):2829, 2021.

\bibitem{MalwareBazaar}
MalwareBazaar.
\newblock Malware sharing platform, 2024.
\newblock Accessed: 2024-05-14.

\bibitem{intro_fsl_ft_1}
Akihiro Nakamura and Tatsuya Harada.
\newblock Revisiting fine-tuning for few-shot learning, 2019.

\bibitem{neyshabur2017exploring}
Behnam Neyshabur, Srinadh Bhojanapalli, David McAllester, and Nathan Srebro.
\newblock Exploring generalization in deep learning, 2017.

\bibitem{rw_cv_2}
H.~Noh, S.~Hong, and B.~Han.
\newblock Learning deconvolution network for semantic segmentation.
\newblock In {\em 2015 IEEE International Conference on Computer Vision (ICCV)}, pages 1520--1528, Los Alamitos, CA, USA, dec 2015. IEEE Computer Society.

\bibitem{chatgpt}
OpenAI.
\newblock {ChatGPT}: A large-scale generative model for conversation.
\newblock \url{https://openai.com/research/chatgpt}, 2022.

\bibitem{rw_cv_1}
Wanli Ouyang, Xingyu Zeng, Xiaogang Wang, Shi Qiu, Ping Luo, Yonglong Tian, Hongsheng Li, Shuo Yang, Zhe Wang, Hongyang Li, Kun Wang, Junjie Yan, Chen-Change Loy, and Xiaoou Tang.
\newblock Deepid-net: Object detection with deformable part based convolutional neural networks.
\newblock {\em IEEE Transactions on Pattern Analysis and Machine Intelligence}, 39(7):1320--1334, 2017.

\bibitem{7808140}
Kuntal~Kumar Pal and K.~S. Sudeep.
\newblock Preprocessing for image classification by convolutional neural networks.
\newblock In {\em 2016 IEEE International Conference on Recent Trends in Electronics, Information \& Communication Technology (RTEICT)}, pages 1778--1781, 2016.

\bibitem{rw_robo_1}
Ali Punjani and Pieter Abbeel.
\newblock Deep learning helicopter dynamics models.
\newblock In {\em 2015 IEEE International Conference on Robotics and Automation (ICRA)}, pages 3223--3230, 2015.

\bibitem{ILSVRC15}
Olga Russakovsky, Jia Deng, Hao Su, Jonathan Krause, Sanjeev Satheesh, Sean Ma, Zhiheng Huang, Andrej Karpathy, Aditya Khosla, Michael Bernstein, Alexander~C. Berg, and Li~Fei-Fei.
\newblock {ImageNet Large Scale Visual Recognition Challenge}.
\newblock {\em International Journal of Computer Vision (IJCV)}, 115(3):211--252, 2015.

\bibitem{sandler2019mobilenetv2}
Mark Sandler, Andrew Howard, Menglong Zhu, Andrey Zhmoginov, and Liang-Chieh Chen.
\newblock Mobilenetv2: Inverted residuals and linear bottlenecks, 2019.

\bibitem{schroff2015facenet}
Florian Schroff, Dmitry Kalenichenko, and James Philbin.
\newblock Facenet: A unified embedding for face recognition and clustering.
\newblock In {\em Proceedings of the IEEE conference on computer vision and pattern recognition}, pages 815--823, 2015.

\bibitem{schürholt2022model}
Konstantin Schürholt, Diyar Taskiran, Boris Knyazev, Xavier~Giró i~Nieto, and Damian Borth.
\newblock Model zoos: A dataset of diverse populations of neural network models, 2022.

\bibitem{intro_osl_cv_2}
Amirreza Shaban, Shray Bansal, Zhen Liu, Irfan Essa, and Byron Boots.
\newblock One-shot learning for semantic segmentation, 2017.

\bibitem{subramanian2021image}
Nandhini Subramanian, Omar Elharrouss, Somaya Al-Maadeed, and Ahmed Bouridane.
\newblock Image steganography: A review of the recent advances.
\newblock {\em IEEE access}, 9:23409--23423, 2021.

\bibitem{szegedy2016inceptionv4}
Christian Szegedy, Sergey Ioffe, Vincent Vanhoucke, and Alex Alemi.
\newblock Inception-v4, inception-resnet and the impact of residual connections on learning, 2016.

\bibitem{szegedy2015rethinking}
Christian Szegedy, Vincent Vanhoucke, Sergey Ioffe, Jonathon Shlens, and Zbigniew Wojna.
\newblock Rethinking the inception architecture for computer vision, 2015.

\bibitem{tan2021efficientnetv2}
Mingxing Tan and Quoc~V. Le.
\newblock Efficientnetv2: Smaller models and faster training, 2021.

\bibitem{tensorflowhub}
TensorFlow.
\newblock {TensorFlow Hub}, 2022.
\newblock Accessed: 2022-12-19.

\bibitem{verma2023insecure}
Aneesh Verma.
\newblock Insecure deserialization detection in python.
\newblock 2023.

\bibitem{wang2019simpleshot}
Yan Wang, Wei-Lun Chao, Kilian~Q. Weinberger, and Laurens van~der Maaten.
\newblock Simpleshot: Revisiting nearest-neighbor classification for few-shot learning, 2019.

\bibitem{wang2021evilmodel}
Z.~Wang, C.~Liu, and X.~Cui.
\newblock Evilmodel: hiding malware inside of neural network models.
\newblock In {\em 2021 IEEE Symposium on Computers and Communications (ISCC)}, pages 1--7. IEEE, 2021.

\bibitem{wang2022evilmodel}
Z.~Wang, C.~Liu, X.~Cui, J.~Yin, and X.~Wang.
\newblock Evilmodel 2.0: bringing neural network models into malware attacks.
\newblock {\em Computers \& Security}, 120:102807, 2022.

\bibitem{wang2021data}
Zihan Wang, Olivia Byrnes, Hu~Wang, Ruoxi Sun, Congbo Ma, Huaming Chen, Qi~Wu, and Minhui Xue.
\newblock Data hiding with deep learning: A survey unifying digital watermarking and steganography.
\newblock {\em arXiv preprint arXiv:2107.09287}, 2021.

\bibitem{Waskom2021}
Michael~L. Waskom.
\newblock seaborn: statistical data visualization.
\newblock {\em Journal of Open Source Software}, 6(60):3021, 2021.

\bibitem{float16_wiki}
{Wikipedia contributors}.
\newblock Half-precision floating-point format --- {Wikipedia}{,} the free encyclopedia.
\newblock \url{https://en.wikipedia.org/w/index.php?title=Half-precision_floating-point_format&oldid=1223724160}, 2024.
\newblock [Online; accessed 20-May-2024].

\bibitem{float32_wiki}
{Wikipedia contributors}.
\newblock Single-precision floating-point format --- {Wikipedia}{,} the free encyclopedia.
\newblock \url{https://en.wikipedia.org/w/index.php?title=Single-precision_floating-point_format&oldid=1220802754}, 2024.
\newblock [Online; accessed 20-May-2024].

\bibitem{rw_robo_2}
Yezhou Yang, Yi~Li, Cornelia Fermüller, and Yiannis Aloimonos.
\newblock Robot learning manipulation action plans by " watching " unconstrained videos from the world wide web.
\newblock 01 2015.

\bibitem{intro_fsl_da_1}
Jing Zhou, Yanan Zheng, Jie Tang, Jian Li, and Zhilin Yang.
\newblock Flipda: Effective and robust data augmentation for few-shot learning, 2022.

\bibitem{zoph2018learning}
Barret Zoph, Vijay Vasudevan, Jonathon Shlens, and Quoc~V. Le.
\newblock Learning transferable architectures for scalable image recognition, 2018.

\end{thebibliography}

% \newpage
\appendices

\section{Background}
\subsection{X-LSB-Attack}
We include pseudo-code to illustrate better the 2 X-LSB-Attack variants described in Section \ref{sec:x_lsb_attack}. In the pseudo-code, we use $w[0:r]$ to denote the first $r$ bits of the float number $w$. $v+u$ denotes the concatenation of 2 binary strings, and $v^c$ denotes concatenating $v$ to itself $c$ times.

\begin{algorithm}
\caption{X - LSB attack}\label{alg:x_lsb_attack}
\hspace*{\algorithmicindent} \textbf{Input:} Integer $1 \le X \le s$,\\
List of float numbers $W:=[w_1, w_2, ..., w_n]$,\\
binary malware $m:=b_1\cdots b_k$ \\
\hspace*{\algorithmicindent} \textbf{Output:} List of float numbers $\widehat{W}$ with $m$ embedded in its numbers.
\begin{algorithmic}[1]
\If{$k > n \cdot X$}
    \State $m$ cannot fully fit inside $W$, terminate.
\EndIf
\State Set $t:=\lceil k/X \rceil$
\State Let $m_i:=b_{(i-1)X+1}\cdots b_{iX}$  $\forall 1 \le i < t$
% \Statex Set $m_{parts}:=[m_1,...,m_{t}]$
\Statex $m_{remainder}=m_t:=b_{(t-1)X+1}\cdots b_k$
% \State Let $bw_i$ a binary representation of $w_i$ $\forall 1 \le i < n$
% \State Let 
\State Set $\widehat{w_i}:=w_i[0:s-X]+m_i$ $\forall 1 \le i < t$
\State Set $\widehat{w_t}:=w_t[0:s-X]+m_t$
\Statex \textbf{Return} $\widehat{W}:=[\widehat{w_1}, ..., \widehat{w_t}, w_{t+1}, ..., w_n]$
\end{algorithmic}
\end{algorithm}

\begin{algorithm}
\caption{X - LSB attack - Fill}\label{alg:x_lsb_attack_fill}
\hspace*{\algorithmicindent} \textbf{Input:} Integer $1 \le X \le s$,\\
List of float numbers $W:=[w_1, w_2, ..., w_n]$,\\
binary malware $m:=b_1\cdots b_k$ \\
\hspace*{\algorithmicindent} \textbf{Output:} List of float numbers $\widehat{W}$ with $m$ fully embedding its X-LSBs.
\begin{algorithmic}[1]

\If{$k > n \cdot X$}
    \State Set $\widehat{m}:=m[0:nX]$
\Else
    \State Set $\widehat{m}:=m^{\lceil nX/k \rceil}[0:nX]$
\EndIf
\Statex \textbf{Return} X-LSB-Attack($X$, $W$, $\widehat{m}$) 
\end{algorithmic}

\end{algorithm}

\subsection{Floating Point Format}
\label{sec:float}
Floating-point numbers \cite{float32_wiki, float16_wiki} (also called floats) represent decimal numbers in computer memory. There are multiple standard types of floats with varying precision, such as single-precision float (also called float32), which takes up 32 bits in memory; half-precision float (also called float16), which takes up 16 bits in memory, and more. Float numbers are divided into three sections: sign, exponent, and mantissa (also called fraction). The sign bit denotes the sign of the number, the exponent section represents a negative or positive power of 2, and the mantissa section represents the precise value after the decimal point. We proceed to explain more about float32, which is the main data type we experiment with.

\subsubsection{Float32}
\begin{figure}[!h]
\centering
\includegraphics[width=\linewidth]{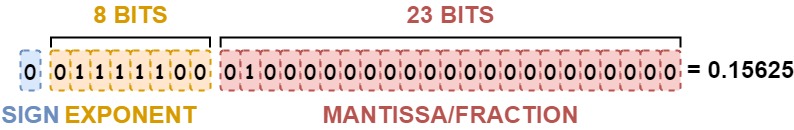}
\caption{Illustration of float32 numbers. In this example, the sign bit $b_{31}$ equals $0_2=0_{10}$, the exponent section $b_{30}b_{29}\cdots b_{23}$ equals $01111100_2=124_{10}$, and the mantissa region $b_{22}b_{21} \cdots b_{0}$ equals $01\cdots0_2$. Hence, the number equals $(-1)^{b_{31}} \cdot 2^{(b_{30}b_{29}\cdots b_{23})_2} \cdot (1.b_{22}b_{21} \cdots b_{0})_2 = (-1)^{0} \cdot 2^{124-127} \cdot (1.01\cdots 0)_2 = +2^{-3}\cdot 1.25 = 0.15625$}
\label{fig:float32}
\end{figure}

Float32 takes up 32 bits in memory, as the name suggests. The 32 bits are segmented into 1/8/23 portions for the sign/exponent/mantissa sections. For some binary float32 number $b_{31}b_{30}\cdots b_{1}b_{0}$ where $b_{31}$ is the sign bit, $b_{30}b_{29}\cdots b_{23}$ is the exponent section and $b_{22}b_{21} \cdots b_{0}$ is the mantissa section, the decimal value of the number is calculated by
\[(-1)^{b_{31}} \cdot 2^{(b_{30}b_{29}\cdots b_{23})_2 - 127_{10}} \cdot (1.b_{22}b_{21} \cdots b_{0})_2\]
See Figure \ref{fig:float32} for an illustration.

\section{AI Model Image Representation}
\begin{figure}[!h]
\centering
\includegraphics[width=\linewidth]{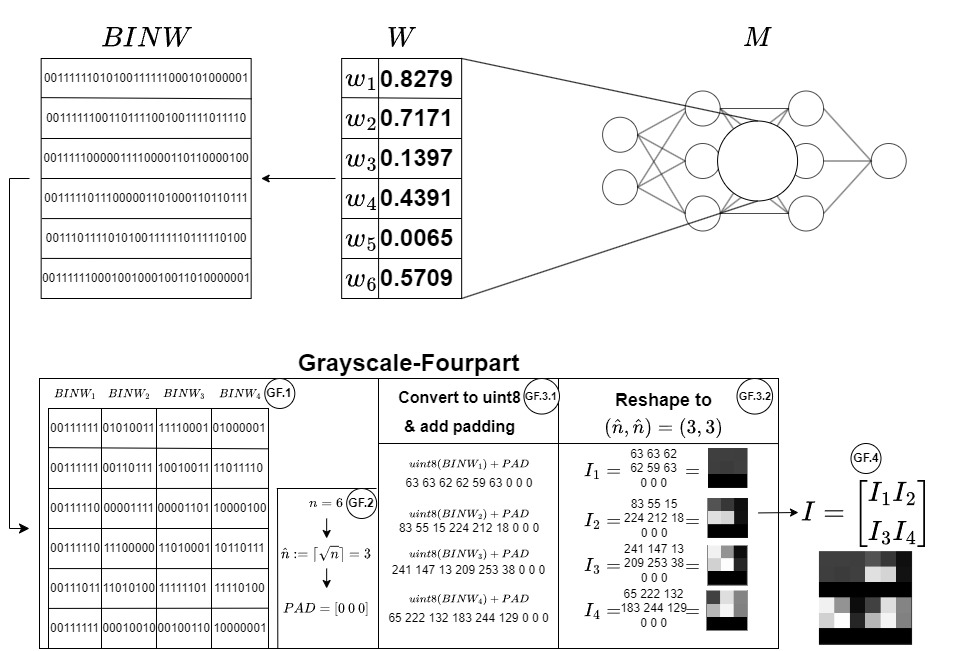}
\caption{Illustration of a technique for creating image representations from binary form float32 data ($BINW$). The technique is used to create images from model weights and gradients. The Grayscale-Fourpart (GF) technique splits every float32 into 4 segments (GF.1) - resulting in 4 vectors of 8-bit binary values ($BINW_1, BINW_2, ...$). To turn the vectors into a square shape, the closest square root from above ($\hat{n})$  is calculated, and zero padding is applied as needed (GF.2). Each byte vector is converted from binary form to uint8 (0-255) and the zero padding is added (GF.3.1). The vectors are reshaped to a square matrix - each acting as an image (GF.3.2), Finally, the Images are stacked to form the output grayscale 8-bit image $I$ (GF.4).}
\label{fig:model2img}
\end{figure}

See Algorithm \ref{alg:grayscale_fourpart} for the pseudo-code of the Grayscale-Fourpart image representation, and see Figure \ref{fig:model2img} for an illustration of the image model representation creation process.

\begin{algorithm}
\caption{Grayscale-Fourpart}\label{alg:grayscale_fourpart}
\hspace*{\algorithmicindent} \textbf{Input:} Vector of float32 data in binary form $BINW$ of shape $(n,32)$ \\
\hspace*{\algorithmicindent} \textbf{Output:} 8-bit Grayscale Image $I$
\begin{algorithmic}[1]
\State Divide $BINW$ into 4 parts (each of shape $(n, 8)$): 
\Statex    $BINW_j:=BINW[:,8(j-1):8j]$ for $1\le j \le 4$
% \STATEx    $BINW_2:=BINW[:,8:16]$, 
% \STATEx    $BINW_3:=BINW[:,16:24]$, 
% \STATEx    $BINW_4:=BINW[:,24:32]$
\State let $\hat{n}:=\lceil \sqrt{n} \rceil$ the nearest square of n (from above)
\Statex let $PAD = $ zero vector of shape $(\hat{n}^2-n,1)$
\State Calculate unsigned int8 matrices (each of shape $(\hat{n}, \hat{n})$):
\Statex    $I_j:=uint8(BINW_j)+PAD$ reshaped to shape $(\hat{n}, \hat{n})$ for $1\le j \le 4$
% \STATEx    $I_2:=uint8(BINW_2)$,
% \STATEx    $I_3:=uint8(BINW_3)$,
% \STATEx    $I_4:=uint8(BINW_4)$,
\State \textbf{Return} $I:=\begin{bmatrix}I_1 I_2\\I_3 I_4\end{bmatrix}$
\end{algorithmic}
\end{algorithm}

\section{Model Performance}
\label{sec:model_performance}
In this section, we present model evaluation results of benign and attacked models from our created datasets (Section \ref{sec:datasets}) to illustrate how the steganography attacks affect the model performance on some validation dataset. The models from the SCZ dataset are each tested on their corresponding dataset, and the models from the famous CNN architectures are tested against the ImageNet2012 \cite{ILSVRC15} validation dataset.
See Table \ref{tab:model_eval_scz}, Table \ref{tab:model_eval_f10m}, and Table \ref{tab:model_eval_f100m}.
We can see that when compared with the benign model results, attacks up to 16 bits (50\% embedding rate) cause almost no degradation in model performance. This means attackers would opt to leverage 1-16 bits of the parameters to avoid causing suspicion in end-users. Nevertheless, even 1-16 bits, which are 3.1\%-50\% of the whole model, allow an attacker an embedding capacity of 3MB-50MB in a 100MB model. In addition, we can see that using 22 and 23 bits in the attack completely eroded the models in all cases. We can directly conjecture that attacks outside the mantissa will be even more noticeable and unviable for attackers since the exponent bits in float32 can cause exponential changes (Section \ref{sec:float}). This can justify focusing only on the mantissa.

\begin{table}[h]
\caption{Benign and attacked average SCZ model performance on the corresponding validation dataset}
  \label{tab:model_eval_scz}
\centering
\begin{tabular}{lccccccc}
\hline
ZOO\textbackslash{}X & 0 (benign)  & 8           & 16          & 20          & 21 & 22 & 23 \\ \hline
CIFAR10              & \textbf{45} & \textbf{45} & \textbf{45} & \textbf{45} & 43 & 39 & 30 \\
MNIST                & \textbf{81} & \textbf{81} & \textbf{81} & \textbf{81} & 80 & 72 & 64 \\
STL10                & \textbf{34} & \textbf{34} & \textbf{34} & \textbf{34} & 33 & 30 & 25 \\
SVHN                 & \textbf{65} & \textbf{65} & \textbf{65} & \textbf{65} & 63 & 56 & 43 \\ \hline
\end{tabular}
\end{table}

\begin{table}[!h]
\caption{Inference accuracy of famous small CNNs before and after steganography attack. Models were tested on the ImageNet2012 validation dataset. We see that attacks that used 1-16 bits had almost no effect.}
  \label{tab:model_eval_f10m}
\centering
\begin{tabular}{lccccccc}
\hline
ZOO\textbackslash{}X & 0 (benign)           & 8                    & 16                   & 20                   & 21                   & 22                   & 23                   \\ \hline
DenseNet121          & \textbf{74}          & \textbf{74}          & \textbf{74}          & 71                   & 55                   & 9                    & 0                    \\
EfficientNetV2B0     & \textbf{78}          & \textbf{78}          & \textbf{78}          & 75                   & 49                   & 2                    & 0                    \\
MobileNet            & \textbf{70}          & \textbf{70}          & \textbf{70}          & 34                   & 1                    & 0                    & 0                    \\
MobileNetV2          & \textbf{71}          & \textbf{71}          & 70                   & 29                   & 0                    & 0                    & 0                    \\
MobileNetV3Small     & \textbf{68}          & \textbf{68}          & 67                   & 5                    & 0                    & 0                    & 0                    \\
MobileNetV3Large     & \textbf{75}          & \textbf{75}          & \textbf{75}          & 35                   & 1                    & 0                    & 0                    \\
NasNetMobile         & \textbf{74}          & \textbf{74}          & \textbf{74}          & 69                   & 30                   & 0                    & 0                    \\ \hline
\end{tabular}
\end{table}

\begin{table}[!h]
\caption{Inference accuracy of famous large CNNs before and after steganography attack. Models were tested on the ImageNet2012 validation dataset.}
\label{tab:model_eval_f100m}
\centering
\begin{tabular}{lccccccc}
\hline
ZOO\textbackslash{}X & 0 (benign)  & 8           & 16          & 20          & 21 & 22 & 23 \\ \hline
% ConvNeXtBase         & \textbf{85} & \textbf{85} & \textbf{85} & 84          & 81 & 19 & 0  \\
% ConvNeXtSmall        & \textbf{82} & \textbf{82} & \textbf{82} & \textbf{82} & 80 & 41 & 0  \\
% ConvNeXtTiny         & \textbf{81} & \textbf{81} & \textbf{81} & \textbf{81} & 78 & 59 & 0  \\
DenseNet169          & \textbf{76} & \textbf{76} & \textbf{76} & 73          & 60 & 3  & 0  \\
DenseNet201          & \textbf{77} & \textbf{77} & \textbf{77} & 73          & 61 & 1  & 0  \\
EfficientNetV2B2     & \textbf{80} & \textbf{80} & \textbf{80} & 77          & 53 & 0  & 0  \\
EfficientNetV2B3     & \textbf{82} & \textbf{82} & \textbf{82} & 79          & 63 & 1  & 0  \\
InceptionResNetV2    & \textbf{80} & \textbf{80} & \textbf{80} & 78          & 68 & 4  & 0  \\
InceptionV3          & \textbf{77} & \textbf{77} & \textbf{77} & 75          & 60 & 4  & 0  \\
ResNet50             & \textbf{74} & \textbf{74} & \textbf{74} & 70          & 55 & 11 & 0  \\
ResNet50V2           & \textbf{70} & \textbf{70} & \textbf{70} & 66          & 38 & 2  & 0  \\
ResNet101            & \textbf{76} & \textbf{76} & \textbf{76} & 73          & 61 & 8  & 0  \\
ResNet101V2          & \textbf{72} & \textbf{72} & \textbf{72} & 66          & 38 & 1  & 0  \\
ResNet152            & \textbf{76} & \textbf{76} & \textbf{76} & 67          & 48 & 1  & 0  \\
ResNet152V2          & \textbf{72} & \textbf{72} & \textbf{72} & 68          & 54 & 5  & 0  \\
Xception             & \textbf{79} & \textbf{79} & \textbf{79} & 74          & 43 & 0  & 0  \\ \hline
\end{tabular}
\end{table}

\end{document}